\documentclass[12pt]{iopart}

\usepackage{graphicx}
\usepackage{dcolumn}
\usepackage{bm}
\usepackage{longtable}
\usepackage{float}

\begin{document}

\title{Heat accumulation and all-optical switching by domain wall motion in Co/Pd superlattices}

\author{F Hoveyda, E Hohenstein and S Smadici}

\address{Department of Physics and Astronomy, University of Louisville, KY 40292, USA}
\ead{serban.smadici@louisville.edu}

\begin{abstract}
All-optical switching by domain wall motion has been observed in Co/Pd superlattices. Heat accumulation is part of the switching process for our experimental conditions. Numerical calculations point to a connection between domain wall motion and in-plane heat diffusion.
\end{abstract}


\maketitle

\section{Introduction}

In ultrafast all-optical switching (AOS) of ferrimagnetic rare earth - transition metal alloys the magnetization is reversed by 180 degrees.~\cite{2007Stanciu} This reversal was examined in detail and both polarization-independent and polarization-dependent switching have been reported~\cite{2012Vahaplar,2012Khorsand}. It is remarkable that partial ultrafast reversal is observed within $1~ps$, much less than typical precession periods. As also observed with electron beam fields~\cite{2004Tudosa}, the macrospin of the Landau-Lifshitz-Gilbert (LLG) equation fragments in the intense applied laser field. However, it reassembles in the opposite direction through an intermediate ferromagnetic state. Models of magnetization ultrafast time dependence apply the Landau-Lifshitz-Bloch (LLB) equations accounting for the variation in macrospin magnitude~\cite{2012Atxitia} or microscopic atomistic calculations with exchange interactions~\cite{2012Ostler,2014Oniciuc}. More recently, AOS was observed in ferromagnetic Co/Pt superlattices~\cite{2014Lambert,2016Hadri}. Models for ferrimagnets rely on specific microscopic exchange interactions. For instance, for a ferrimagnet with two ($A$ and $B$) sub-lattices, Heisenberg exchange interactions $J_{AA}$, $J_{AB}$, and $J_{BB}$ are considered. Little of this applies to a ferromagnet, which has only one interaction constant $J$, leaving open the question of the origin of AOS in these materials.

Previous experiments on ferrimagnetic materials were done with single-pulse excitation. The relatively low repetition rate (kHz range) lasers used in these studies allowed sufficient movement of the sample between pulses, so that pulses did not overlap over the same sample area. This is important in ferrimagnets because each pulse initiates a reversal~\cite{2012Ostler}. The AOS in ferromagnetic Co/Pt also utilized a low-repetition rate laser. However, AOS was observed even when the pulses overlapped on the sample. This suggested that it may be possible to initiate magnetization reversal in ferromagnets with high-repetition rate (tens of MHz range) lasers.

In this work, we scanned an ultrafast high-repetition rate laser field on Co/Pd superlattices. Polarization-independent AOS was observed. In our experimental conditions, reversal by domain wall (DW) motion becomes visible after a few $ms$. Polarizing and magnetic microscopy images show a reversal driven by heat accumulation in the sample and in-plane thermal gradients.

\section{Experiments}

\subsection{Samples and experimental setups}

The [Co/Pd] multilayer samples $A$ and $B$ were grown by e-beam evaporation at room temperature. Before the deposition, the Corning white-water glass substrates were immersed in Nanostrip solution for five minutes, placed in acetone and then methanol, and sonicated in each liquid for 10 minutes at $60~^{o}C$. The deposition was carried out at $2 \times 10^{-6}$ Torr pressure with the substrate at an angle of $45^{o}$ to the incident beam. The substrate rotated around its normal at $5~ RPM$ during deposition.  The target multilayer structure was 4 x [Co/Pd] with 0.7 nm thick Co and 1 nm thick Pd layers. The total thicknesses measured with AFM were 4.1 nm and 6.2 nm for samples $A$ and $B$. Sample $A$ is thinner because it was placed further out from the crucibles during deposition.

MOKE measurements were performed in both longitudinal and polar geometry. The variable-temperature setup consisted of a HeNe laser, rotatable Glan-Thompson polarizer and analyzer, photoelastic modulator (PEM), an electromagnet (GMW) driven by a bipolar power supply (KEPCO BOP 50-8ML), and a photodiode detector connected to a lock-in amplifier. The sample was mounted on an Al holder and placed at the center of the electromagnet. The incident beam was $p$-polarized and was focused on the sample. The incidence angle in the L-MOKE geometry was 25 deg. The reflected beam from sample surface passed through the PEM (Hinds PEM-90) and analyzer. Hysteresis loops in the P- and L-MOKE geometries were obtained by averaging the results from running a sixteen-cycle waveform (figure 1(a)).

\begin{figure}
\centering\rotatebox{0}{\includegraphics[scale=0.65]{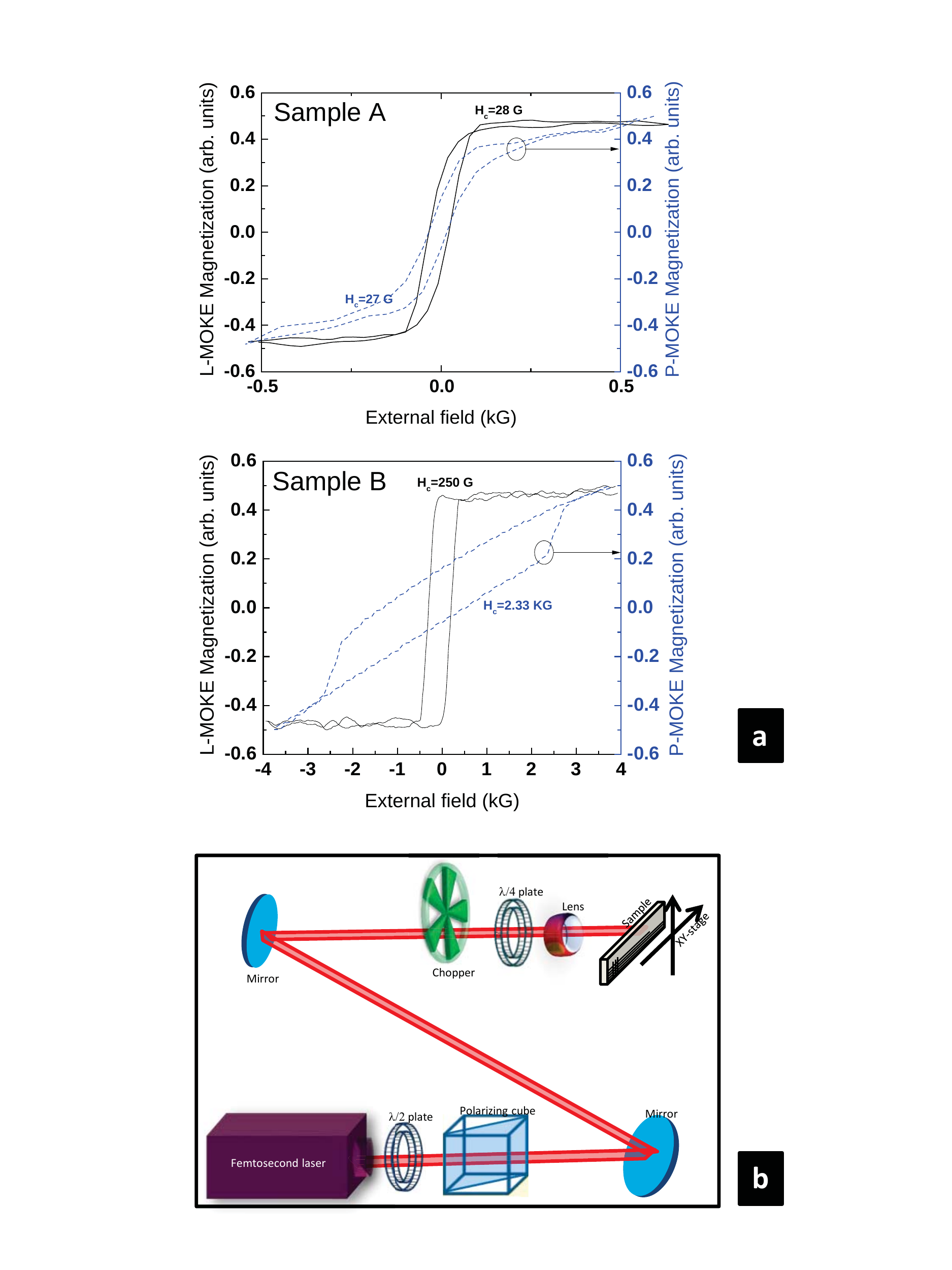}}
\caption{\label{fig:Figure1} (a) L- and P-MOKE hysteresis loops at room temperature for samples $A$ and $B$. (b) Sketch of the writing setup.}
\end{figure}

The fluence of an ultrafast TiS laser with $120~fs$ pulses at $800~nm$ wavelength and a repetition rate of $80~MHz$ was adjusted by a combination of a polarizing cube and a half-wave plate (HWP) attached to a motorized rotation stage (figure 1(b)). A quarter-wave plate (QWP) was applied to modify the beam polarization. The laser beam was focused to a typical size of $25~\mu m$. The sample was AC demagnetized before the writing, to increase the number of magnetic domains, and was scanned along the two directions orthogonal to the beam using two motorized stages. The writing speed $v_{s}$ was varied between $0.1~mm/s$ and $10~mm/s$. A small beam asymmetry could be removed by placing a blade before the lens to partially cut the beam.

The sample response to TiS laser scans was examined with polarizing and magnetic force microscopy (MFM). Polarizing microscopy imaging in transmission mode was carried out with a Zeiss Imager microscope. The polarizer and rotatable analyzer were crossed, then offset a small angle in both directions. The images were then subtracted to enhance the birefringent contrast. Images were also made with the polarizer and analyzer removed.

Magnetic domains were imaged using an Asylum Research MFM (MFP-3D-BIO) with a low moment tip from Nanosensors (SSS-MFMR). It was operated in tapping mode at a lift height of 50 nm. The tip was magnetized before imaging with a permanent magnet attached to a moving stage.

\begin{figure}
\centering\rotatebox{0}{\includegraphics[scale=0.75]{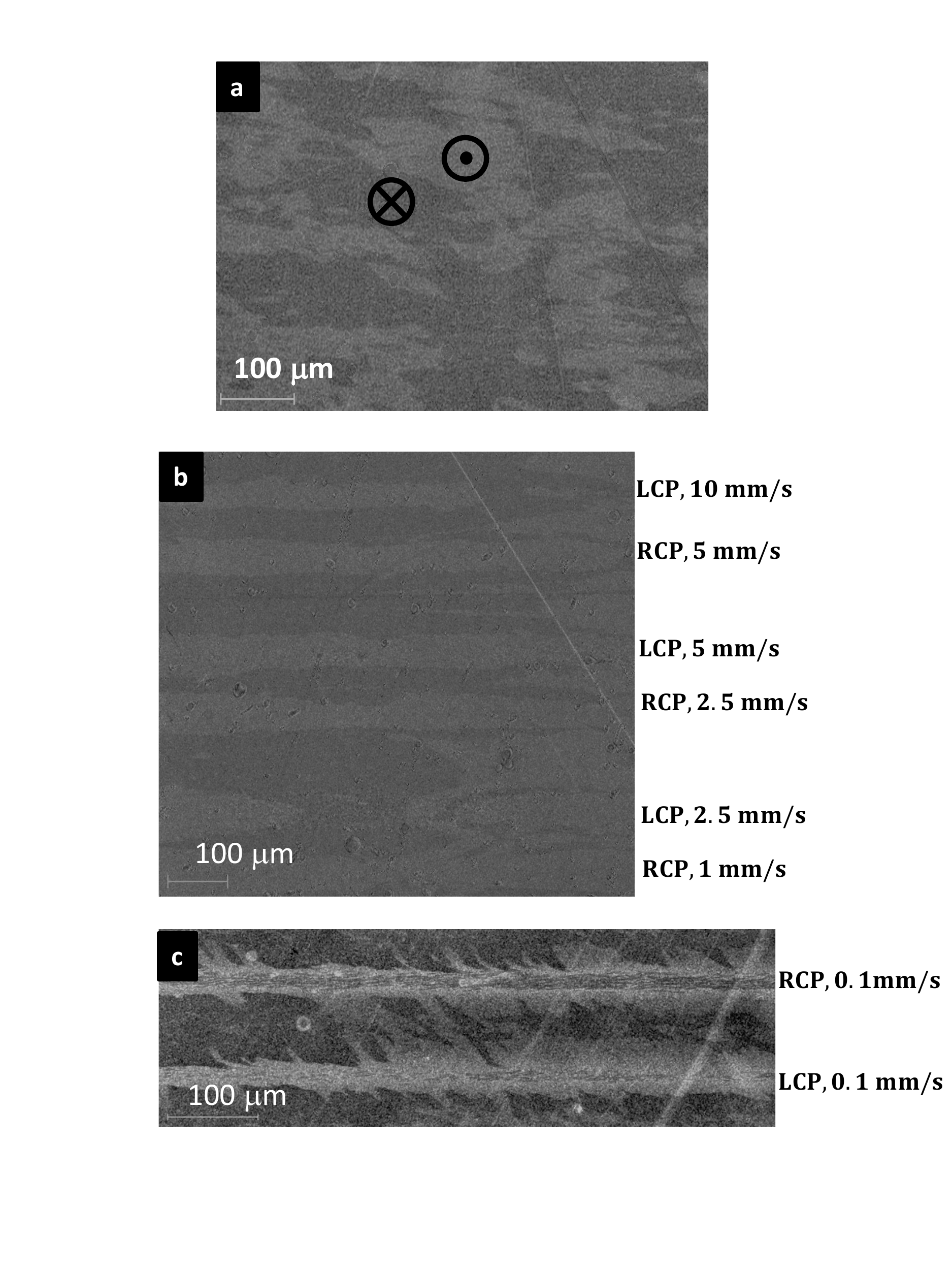}}
\caption{\label{fig:Figure1} (a) Background domains of sample $A$. (b) AOS in sample $A$ at $150~mW$ power. (c) Nucleation of secondary domains at stripe center in sample $A$ at $125~mW$ power and smaller speed.}
\end{figure}

\subsection{Results}

\begin{figure}
\centering\rotatebox{0}{\includegraphics[scale=0.6]{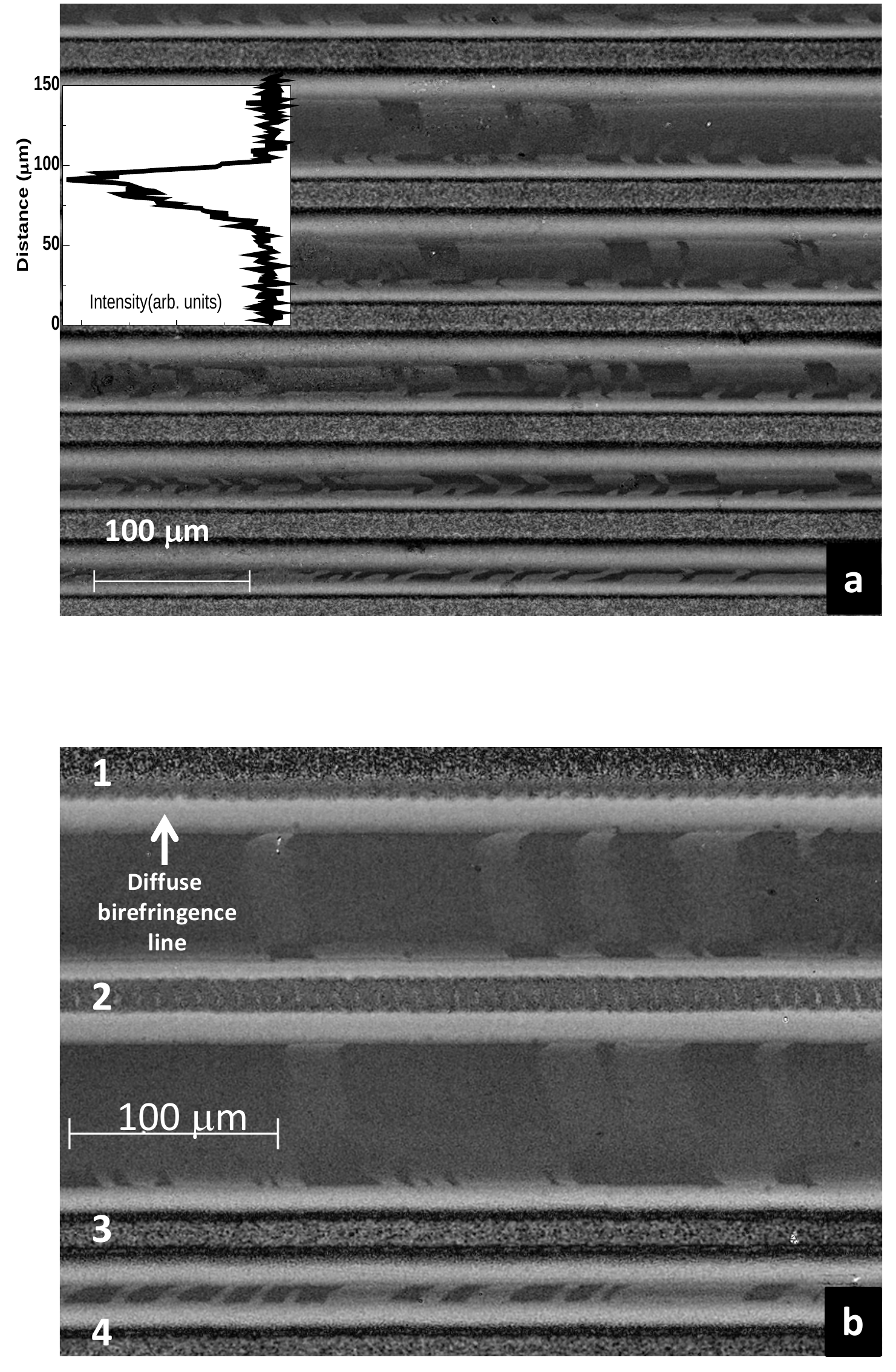}}
\caption{\label{fig:Figure1} Polarizing microscopy images of sample $B$. (a) AOS domains extend away from under the laser beam profile (top) and show complementary contrast at small stripe spacing (bottom). All stripes were written with $600~mW$ power, LP light, and at $10~mm/s$ speed. (b) Opposite reversal across a stripe. Writing parameters are ($380~mW$, LCP, $1~mm/s$) for 1, ($220~mW$, LCP, $1~mm/s$) for 2, ($600~mW$, LP, $10~mm/s$) for 3 and 4.}
\end{figure}

\begin{figure}
\centering\rotatebox{0}{\includegraphics[scale=0.4]{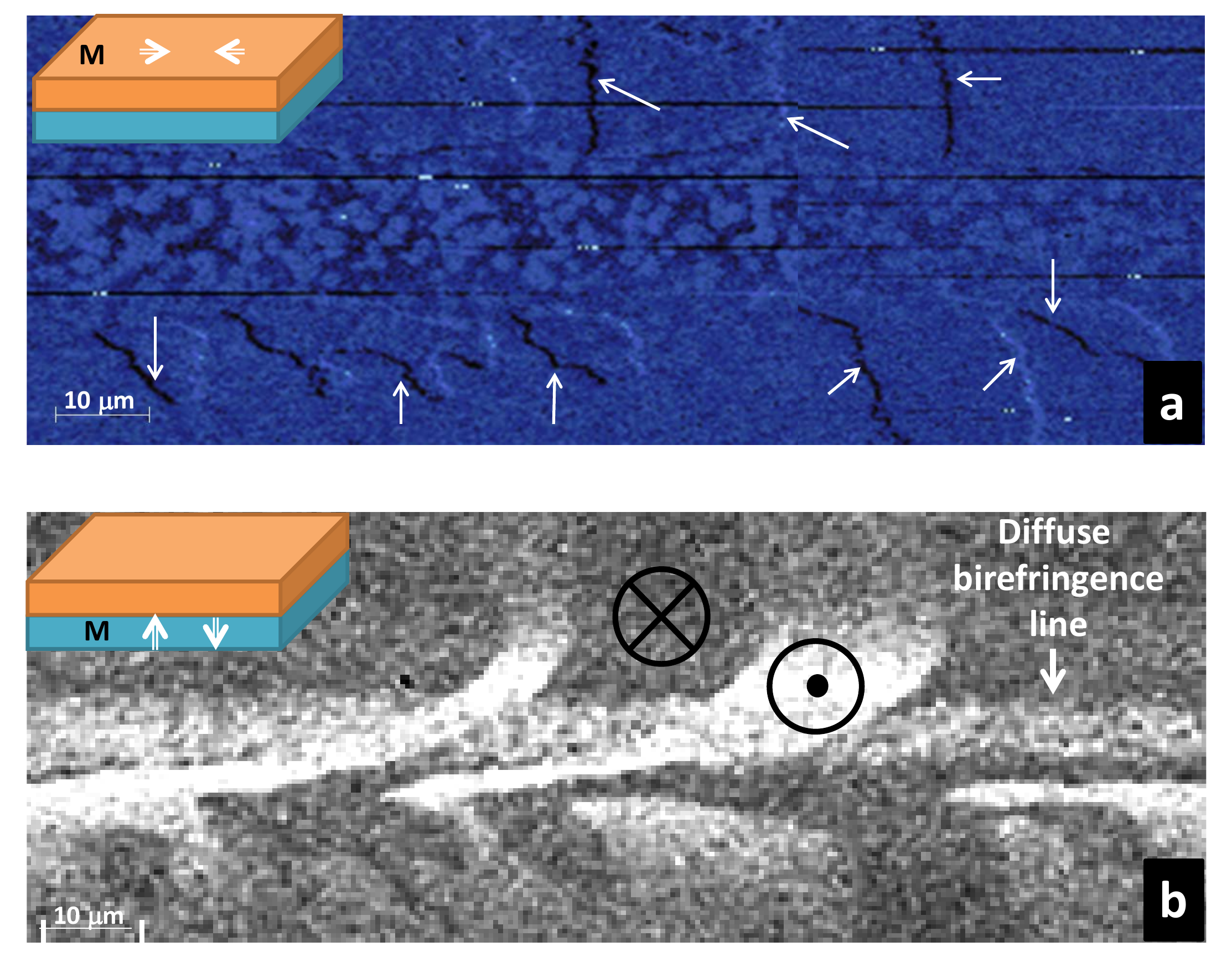}}
\caption{\label{fig:Figure1} MFM (a) and polarizing microscopy (b) images of a stripe in sample $B$ ($100~mW$, LP, $10~mm/s$). The arrows show the domain walls. Inset: sketch of a side view of the magnetic structure.}
\end{figure}

Polarizing microscopy images made before writing show irregular angular magnetic domains (figure 2(a)). Patterns made during the writing procedure appear as ``stripes" in the images. A birefringent contrast at the stripe center is visible after writing (figure 2(b)). This contrast is magnetic, not structural, because it matches the background magnetic domain contrast and angular shape. The contrast is not changed with azimuthal sample rotation, pointing to reversed magnetic domains with magnetization oriented perpendicular to the surface. Reversals with the magnetization pointing down on a magnetization-up background were also observed (not shown).

Further increases in fluence give a thermal demagnetization speckle pattern at stripe center. Regular thermal demagnetization patterns have been obtained before in ferromagnetic Co/Pd superlattices with $12~ns$ pulses ~\cite{2015Stark}. The low-fluence features are no longer visible in the thermally demagnetized regions, unlike what would be expected of structural changes, further supporting their magnetic origin.

Two types of AOS have so far been observed in ferromagnetic materials: single- and multiple-pulse (``cumulative") switching. At a typical writing speed $v_{s}=5~mm/s$ and beam diameter of $25~\mu m$, $4\times 10^{5}$ consecutive pulses are incident on the sample before the beam moves away. Therefore, the observed AOS in Co/Pd is cumulative. Similar results are obtained with right- (RCP), left- (LCP) and linearly-polarized (LP) light, supporting a thermal model for cumulative switching in Co/Pd superlattices. This is in contrast to cumulative switching in Co/Pt superlattices, in which the magnetization was reversed with RCP or LCP light and absent with LP light~\cite{2016Hadri-b}. In addition, different results are obtained with the same pulse power and different scanning speeds, further supporting a thermal model.

Observations also point toward the relevance of long-range processes. Images were made after writing with beams of higher fluence. Fluence decreases gradually toward the wings of the beam profile, from a maximum at the center (figure 3(a), inset). The same sequence of final states is expected across one stripe, as in images of a stripe center made at different powers. Similar results were observed. For instance, the narrowest magnetic domains near the stripes in figure 3(a) are reversed regions, corresponding to regions at the center of the stripe in figure 2. However, for widely-spaced stripes, images also show longer range domains, extending far away from areas exposed to the beam (figure 3(a), top part). Complementary domain contrast is also observed for closely-spaced stripes (figure 3(a), lower part) and for domains on opposite sides of one stripe (figure 3(b)). This suggests that the long-range demagnetization fields should be considered.

MFM measurements gave further insight into the magnetic structure. Small domains are oriented with magnetization perpendicular to the surface at stripe center (figure 4), consistent with previous results~\cite{2015Stark}. Domain walls were observed toward the edges, of the same shape as the domain walls in polarizing microscopy. This shows that the magnetic structure is not uniform within the sample (figure 4, insets), with surface magnetic closure domains forming to minimize the demagnetization energy. In our case, the top layer in-plane magnetization is probed in MFM, but not in polarizing microscopy at normal incidence, while a buried layer out-of-plane magnetization, not visible in MFM, appears in transmission polarizing microscopy.

The orientation of the domain walls (DW) in the immediate vicinity of the stripe makes an ``arrow" pointing in the direction of the laser spot motion on the sample (figure 5). This suggests that DW processes should be considered in a reversal model. DW may change direction further away from the stripe as in figure 3(b), from a residual in-plane magnetic anisotropy induced during deposition.

A diffuse bright birefringent contrast near the thermal demagnetization areas, without the angular edges of magnetic domains, is also visible in polarizing microscopy images (figure 3). Unlike the magnetic domains, this birefringence contrast changes sign with sample rotation with a period consistent with structural birefringence of the glass substrate ~\cite{2001Sudrie,2004Yang} and small linear defects induced by the laser beam ~\cite{2006Bhardwaj,2009Cheng}. This feature is absent in MFM images (figure 4(a)), but present in the polarizing microscopy image (figure 4(b)), also consistent with a substrate location.

The importance of DW points to an expanded timescale of the switching process. To investigate this more directly, measurements were made with a chopped beam and moving samples. This resulted in sequences of reversed ``dots", each corresponding to the time intervals during which the chopper blade did not block the beam (figure 6). Domains show a filamentary structure in polarizing microscopy images, as they grow from under the laser beam. The front-back asymmetry, with domains offset in the scanning direction with respect to the diffuse background, shows a delay on the order of $\sim \frac{25 ~\mu m}{5 ~ mm/s} = 5~ms$ in the initiation of magnetization switching. Therefore, a slow process is part of the complete reversal. Fully developing AOS in Co/Pd takes significant time.

\begin{figure}
\centering\rotatebox{0}{\includegraphics[scale=0.45]{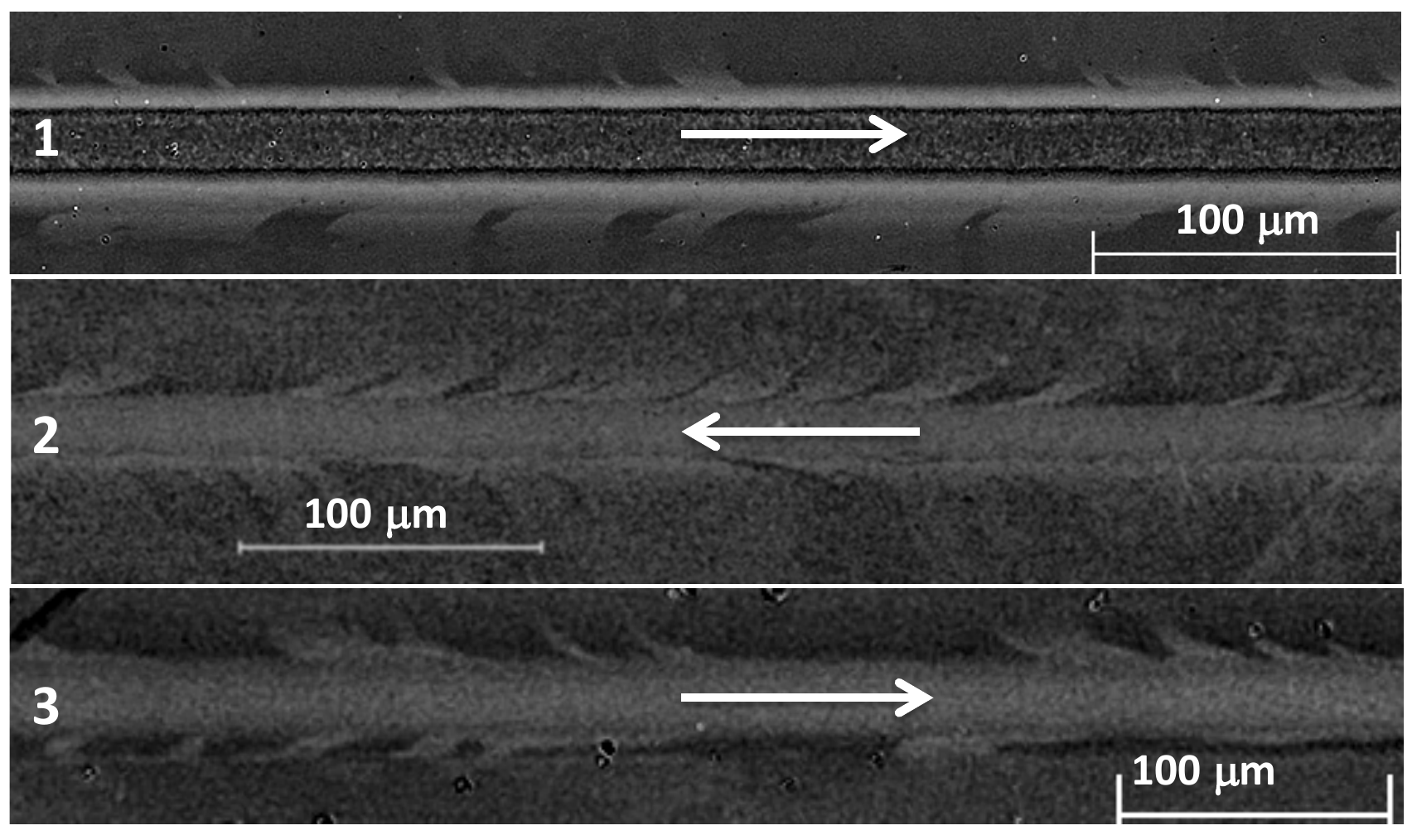}}
\caption{\label{fig:Figure1} Domain walls are oriented differently, depending on the scan direction. The arrows show the direction of laser beam motion. Writing parameters for stripe 1, 2, and 3 were (sample B, $600~mW$, LCP, $10~mm/s$), (sample $A$, $125~mW$, RCP, $10~mm/s$), and (sample $A$, $190~mW$, RCP, $10~mm/s$), respectively.}
\end{figure}

\begin{figure}
\centering\rotatebox{0}{\includegraphics[scale=0.6]{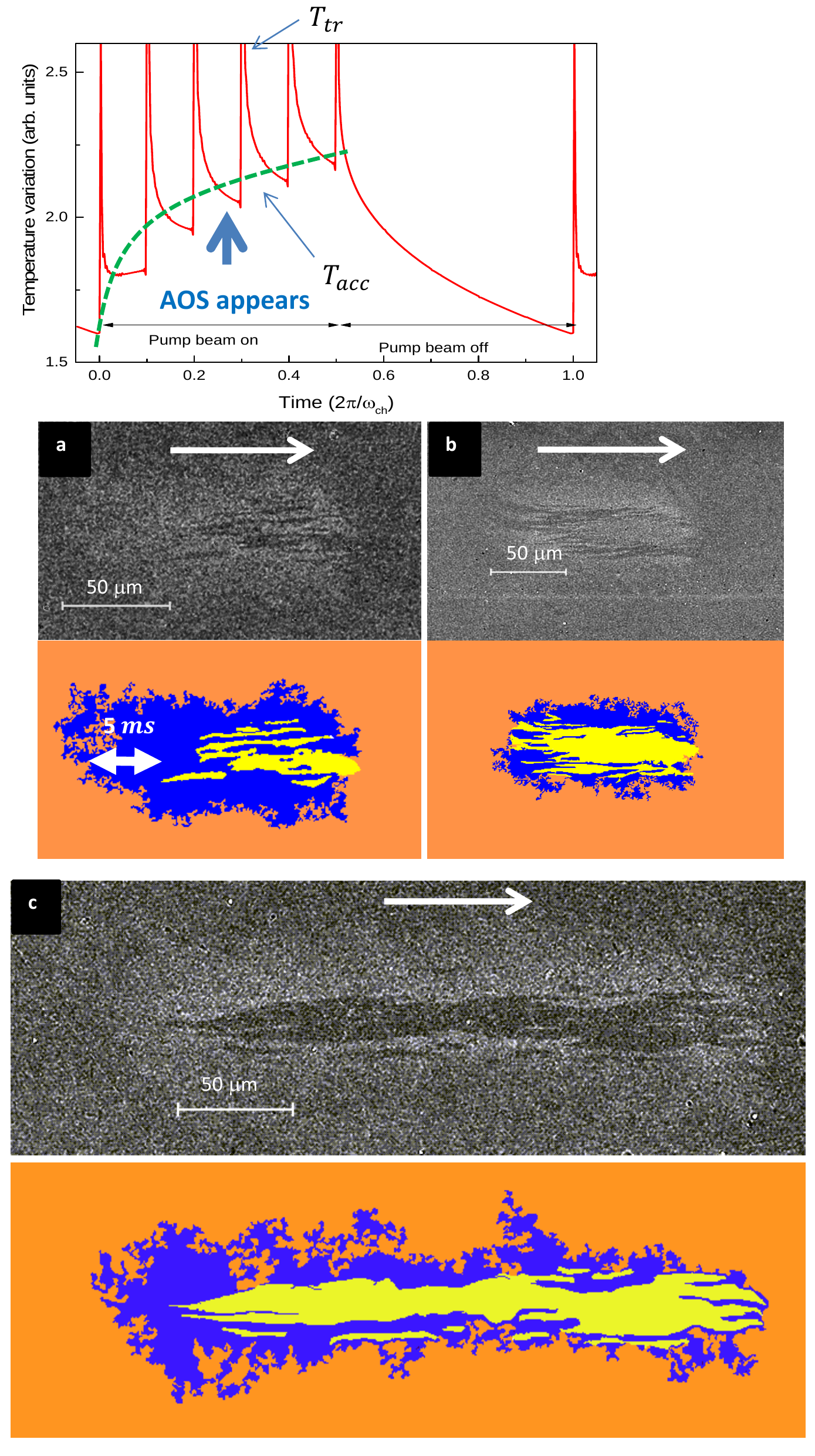}}
\caption{\label{fig:Figure1} Magnetic dots made with chopped beams show a delay in the onset of reversal. These stripes were written in sample $A$. Writing parameters for (a), (b) and (c) were ($f=20~Hz$, $230~mW$, LCP, $5~mm/s$), ($f=10~Hz$, $230~mW$, LCP, $2.5~mm/s$) and ($f=10~Hz$, $230~mW$, RCP, $10~mm/s$), respectively, where $f$ is the chopping frequency. The arrows show the direction of laser beam motion. The top panel shows the heat accumulation at the center of a Gaussian beam.}
\end{figure}

\section{Discussion}

The high-repetition rate laser requires a reconsideration of the assumptions made in previous experiments with low-repetition rate sources. Stronger thermal effects are expected compared to previous AOS experiments.

The long timescale filamentary DW structure in dot images supports a reversal by DW motion in our case, with magnetization rotation at the DW location. Models of ultrafast reversal with uniform magnetization rotation are replaced with a model of reversal by DW motion.

\subsection{Heat accumulation and in-plane heat diffusion}

The laser outputs a sequence of pulses. At the shortest timescale (up to a few ps after the pulse), the sharp transient increase in temperature $T_{tr}(t)$ of the lattice is illustrated by the red peaks in the sketch in the top panel of figure 6, where the $12.5~ns$ spacing between the peaks has been greatly increased for clarity. There are also transient electron $T_{e}$ and spin $T_{s}$ temperatures in three-temperature (3T) models of magnetic materials.~\cite{2005Koopmans,2014Gunther,2014Kuiper}. $T_{e}$ and $T_{s}$ have strong variations over a few ps, after which all three temperatures are the same $T=T_{p}=T_{s}=T_{e}$, gradually decreasing toward the initial temperature. In-plane heat diffusion can be neglected at these timescales, with the models becoming effectively one-dimensional.

At the intermediate timescale, in-plane heat diffusion, which is relatively slow and can be safely neglected in the shortest range, should be included. The heat diffusion equation for an material isotropic in the surface plane is
\begin{eqnarray}
\rho c \frac{\partial T}{\partial t}=K_{||} \Bigg(\frac{\partial^2 T}{\partial x^2 }+\frac{\partial^2 T}{\partial y^2 }\Bigg)+K_{z}\Bigg(\frac{\partial^2 T}{\partial z^2 }\Bigg)
\end{eqnarray}
\noindent where $\rho,c,K_{||},K_{z}$ are the density, the specific heat, in-plane and along $z-$axis thermal conductivities of the material. Heat diffusion occurs initially mostly in the film ($K_{||} \gg K_{z}$). Even with the large diffusivity $D=\frac{K}{\rho c}$ of bulk metals, it would take $t\sim \frac{d^{2}}{D}\sim \frac{(30~\mu m)^{2}}{0.1~cm^{2}/s}\sim 0.1 ~ms \gg t_{rep}=12.5~ns=\frac{1}{\nu_{rep}}$ for heat to diffuse over a distance equal to the beam diameter. The high repetition rate leaves too little time for the sample to cool to the initial temperature after each pulse and heat accumulation between pulses cannot be neglected.~\cite{2008Schmidt} However, the film does not cool very effectively because it is very thin. Therefore, at the longest timescale heat diffusion into the substrate becomes increasingly more important, making up for a reduced $K_{z}$ with a larger heat transfer area. Although the film is the primary factor in absorbing the light and heating the sample, the time evolution of the temperature during cooling is determined mainly by the substrate.

These processes combine to give a gradual heat accumulation and rise in sample temperature. The solution $T(t)$ is known in closed form for a uniform beam profile~\cite{1953Jaeger}. For a stationary beam with a Gaussian profile, the heat accumulation and cooling can be directly illustrated when the small sideways heat transfer in the film is neglected. The time dependence of the temperature at the surface of a semi-infinite medium is obtained by solving the heat diffusion equation, following excitation by a sequence of $\delta-$ function pulses, periodically interrupted by the chopper. It is advantageous to write the temperature as a sum over different frequency components
\begin{eqnarray}
T(t,r)=\rm{Re}  \sum_{l=0}^{\infty}\Bigg[ \frac{\pi}{4}T_{0}(t,r,l\omega_{rep})+\sum_{m=0}^{\infty}\frac{1}{2i(2m+1)}     \times      \\ \nonumber
\Bigg( T_{0}\Big(t,r,l\omega_{rep} +(2m+1)\omega_{ch} \Big)-T_{0}\Big(t,r,l\omega_{rep} -(2m+1)\omega_{ch}\Big) \Bigg) \Bigg]
\end{eqnarray}
\noindent where the sums over $l$ and $m$ are over the femtosecond comb and the chopper square-wave spectrum, and solve the heat diffusion equation at each frequency separately. Then, $T_{0}(t,r,\omega)=\int_{0}^{\infty} dk k J_{0}(kr) \frac{1}{\sqrt{k^{2}+\frac{ i \omega}{D} }}e^{-\frac{k^{2}w_{0}^2}{8}} e^{i\omega t}$, where  $w_{0}$ is the beam diameter and $D$ is the heat diffusion constant. The temperature $T(t,r=0)$ at the center of the beam is plotted in figure 6 and shows the gradual heat accumulation $T_{acc}(t)$ (green dashed line in the top panel) for $\frac{DT_{ch}}{w_{0}^{2}}=0.04$, where $T_{ch}$ is the chopping period. For clarity $\frac{T_{ch}}{T_{rep}}=10$ has been chosen, or 10 pulses incident on the sample during the chopping period (in practice this number is $\sim 10^{5}$).

This time-dependence is being modified when a neighboring point on the sample is considered because the attenuation length from the heat diffusion equation $\lambda_{att}=\sqrt{2D/\omega}$ depends on $\omega$. The $T_{tr}$ spikes, with a large $\omega$, are damped out in $<1 \mu m$. However, the more slowly-changing $T_{acc}$ extends over a considerable distance ($\sim 250~\mu m$ in the film and $\sim 25 ~\mu m$ in glass). With time, heat diffusion appears as a strongly-damped ``wave" with large thermal gradients $\partial T/ \partial r$ along the surface at locations away from the initial disturbance.

A quantitative analysis of heat accumulation and diffusion and a solution for $T_{acc}(x,y,z,t)$ when the beam is moved on the sample is a more complicated problem because the time it takes a moving beam to traverse a distance equal to its diameter $t\sim \frac{25~\mu m}{10~ mm/s} = 2.5~ ms$ at typical scanning speeds $v_{s}=10~mm/s$ is comparable to the heat diffusion time. A numerical solution of the heat diffusion equation for moving beams is required and is presented in the next section.

\subsection{Domain wall dynamics}

The gradual heat accumulation explains the delay in the onset of AOS in figure 6. The chopper turns the pump laser beam on for half of the chopping period. During the ``on" interval, the sequence of pulses gradually increases the temperature $T_{acc}$, shown by the dashed green curve. If we consider a stationary beam first, the sample partly cools during the ``off" interval, and the process is repeated, with the temperature rising until a dynamic equilibrium is obtained between a periodic heating and cooling of the sample, modulated at the chopper frequency. This temperature modulation has been confirmed in our samples in separate pump-probe experiments (in preparation). When we scan the sample under the chopped beam, areas exposed first will have a smaller temperature increase because the beam immediately moves away. In contrast, later areas will receive a larger number of pulses as the beam sweeps across them. This explains the AOS delay in figure 6, when energy has to be deposited first in the structure during the first few $ms$ for domains to become visible, either because AOS is not initiated or because DW do not move well unless $T_{acc}$ is sufficiently large.

DW motion at increased temperatures induced by laser fields has been investigated extensively in magnetic bubble materials. The energy of a cylindrical magnetic domain of radius $r$ and thickness $h$ with magnetization $M_{s}$ perpendicular to the surface in the limit $r \gg h$ is
\begin{eqnarray}
E=2\pi r h \sigma_{w} + 4\pi r h^{2}M_{s}^{2} -8 \pi r h^{2}  {M_{s}}^{2}  \ln \Big(\frac{8r} {h}\Big) +2 \pi r^{2}h M_{s} H
\end{eqnarray}

\noindent where $\sigma_w$ is the domain wall energy and $H$ is the applied external magnetic field. The first term is the domain wall energy, the second and third are the demagnetization field energy ~\cite{1969Thiele,1971Thiele-a,1971Thiele-b}. The DW equilibrium condition can be expressed as a balance of three forces which, when $\sigma_{w}$ and $M_{s}$ are constant, are given by the derivatives of the corresponding energy terms: $F_{w}=-2\pi h\sigma_{w}$ for the domain wall energy, $F_{D}=4\pi h^{2} M_{s}^{2}\Bigg(1+2 \rm{ln} \Big(\frac{8r}{h}\Big) \Bigg)$ for the demagnetization field energy and $F_{H}=-4\pi r h M_{s} H$
for the external field energy ~\cite{1969Thiele,1971Thiele-a,1971Thiele-b}. $F_{w}$ tends to decrease the domain size, while $F_{D}$ tends to increase it. The direction of $F_{H}$ depends on the direction of the external field. The condition $F_{w}+F_{D}+F_{H}=0$ can give a stable solution, if the external field $H$ is not too large, which has been the subject of extensive investigations in magnetic bubble materials, for instance in iron garnet ~\cite{1972Ashkin,1986Kaneko}.

We apply the same approach to explain our observations. In contrast to experiments in magnetic bubble materials, no external field $H$ is applied in our case (the light magnetic field is negligible at our fluence) and $F_{H}=0$. The superlattices are also much thinner, to obtain the required perpendicular magnetic anisotropy. However, the domain wall and demagnetization energy remain comparable because $M_{s}$ is larger than in magnetic bubble materials, as has been confirmed for Co/Pt~\cite{2016Hadri}. The importance of demagnetization energy is further supported by the observed surface magnetic closure domains. The equation $F_{w}+F_{D}=0$ has one solution, which corresponds to the unstable solution of magnetic bubble materials (the stable solution is at infinite radius). It is unstable because increasing $r$ slightly, increases the outward force $F_{D}$, further pushing the DW to larger values. Similarly, decreasing $r$ slightly reduces $F_{D}$ and the domain collapses. In practice, this solution is stabilized because, to move a DW in a sample with a finite coercive field $H_{c}$, it is necessary to supply energy to compensate for energy dissipation associated with DW motion. This gives in the steady-state ($v=const.$)
\begin{eqnarray}
v=\frac{\eta}{2M_{s}}\Bigg( \frac{F}{2\pi  r h} \Bigg )- \eta H_{c}
\end{eqnarray}
\noindent where the case of domain expansion is considered ($v>0$), $\eta$ is the DW mobility and the net force $F$ has been divided by the DW area~\cite{1971Thiele-b,1971Callen}. The last term changes sign when the domain contracts ($v<0$).

The force that overcomes the coercive field term in the velocity equation above and pushes the DW from under the laser beam comes from heat accumulation ($F_{u}$, corresponding to time variations of a spatially uniform temperature $T$) and temperature in-plane gradients ($F_{g}$). The force $F_{u}$ for temperature changes from $T$ to $T'$ is
\begin{eqnarray}
\frac{F_{u}}{2\pi rh}=\frac{\Delta (F_{w}+F_{D})}{2\pi rh}=-\frac{\Delta \sigma_{w}}{r} + \frac{4 h M_{s}}{r} \Bigg(1+2 \rm{ln} \Big(\frac{8r}{h}\Big) \Bigg) \Delta M_{s}
\end{eqnarray}
\noindent where $\Delta \sigma_{w}=\sigma_{w}(T')-\sigma_{w}(T)$ and $\Delta M_{s}=M_{s}(T')-M_{s}(T)$. For instance, if $\sigma_w$ decreases faster with $T$ than ${M_s}^{2}$ the wall energy is relatively smaller at higher temperatures, $F_{u}$ points outward and the domain expands at high $T$. The opposite situation can also occur~\cite{1969Rossol}. In addition, temperature-induced spatial gradients in $\sigma_w$ and $M_s$ and therefore in the energy $E$ give a force $F_{g}$ equal to~\cite{1971Thiele-a}
\begin{eqnarray}
\frac{F_{g}}{2\pi r h}= -\nabla \sigma_{w}+ 4 h M_{s}\Bigg( -1 + 2 \rm{ln}\Big(\frac{8r}{h}\Big) \Bigg) \nabla M_{s}
\end{eqnarray}
\noindent where $\nabla \sigma_{w}=\frac{\partial \sigma_{w}}{\partial T} \nabla T$ and $\nabla M_{s}=\frac{\partial M_{s}}{\partial T} \nabla T$ are in-plane gradients of DW energy and magnetization induced by an in-plane temperature gradient $\nabla T$. A positive sign of $F_{g}$ corresponds to a force pointing outward in a gradient $\nabla T<0$ (a reduction of $T$ when moving away from the center of the beam). A positive first term requires $\frac{\partial \sigma_{w}}{\partial T} >0$, which, although not the usual situation, has been observed in $\rm Sm_{x}Tb_{1-x}FeO_{3}$~\cite{1969Rossol}. The second term in $F_{g}$ is positive for $\nabla T<0$ and $\frac{\partial M_{s}}{\partial T}<0$ except at very small $r$.

\begin{figure}
\centering\rotatebox{0}{\includegraphics[scale=0.6]{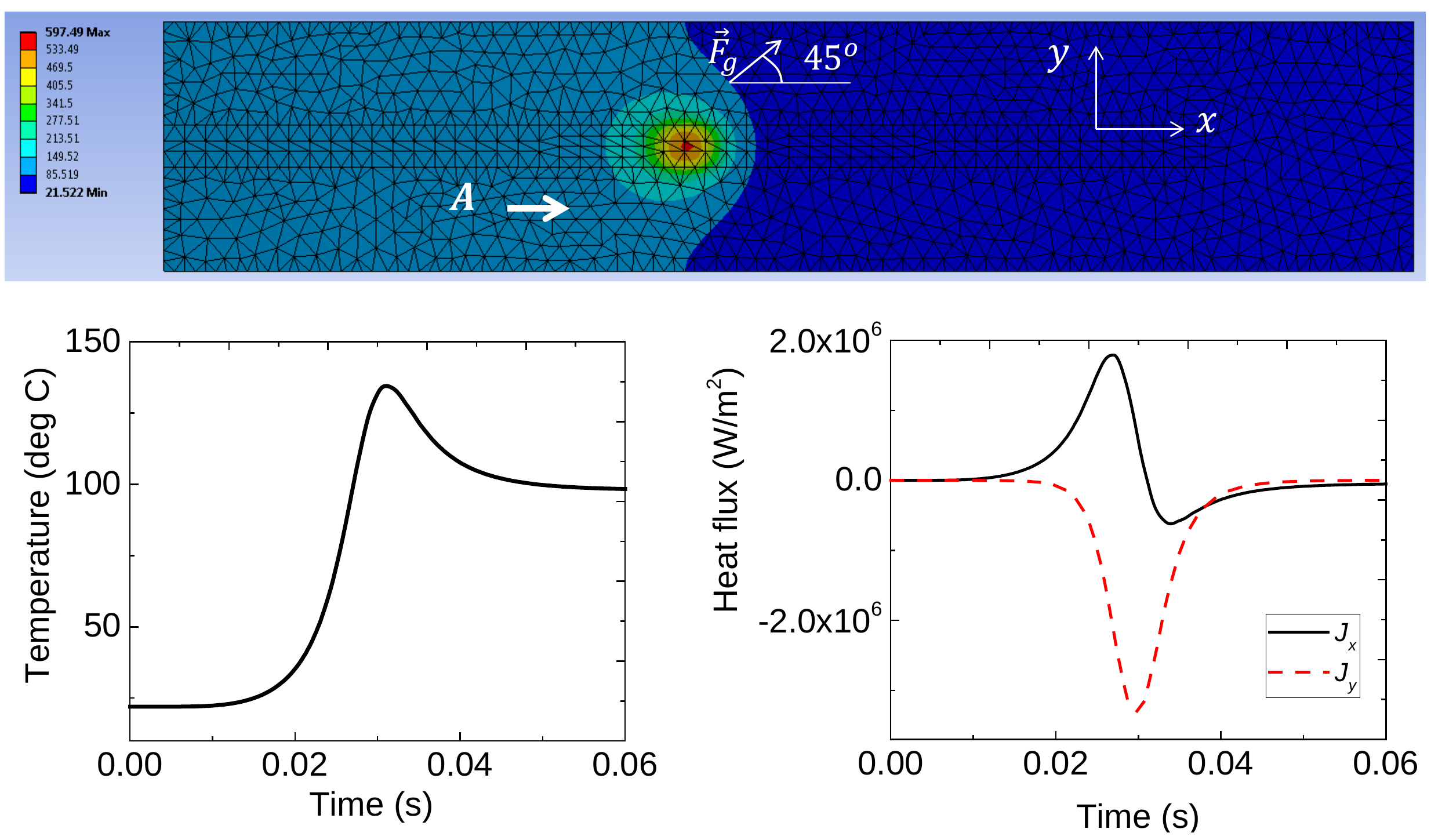}}
\caption{\label{fig:Figure1} Frame at $t=37~ms$ from results of numerical calculations of the temperature variations of the glass substrate with $D_{glass}=0.005~\frac{cm^{2}}{s}$, $v_{scanning}=1~\frac{cm}{s}$, for a $F=5\times10^{7}~\frac{W}{m^{2}}$ fluence focused in an area $A=30~\mu m \times 30~\mu m$. The colour scale is in degrees C. The panels show the time-dependence of the temperature and in-plane heat flux at point $A$ on the surface during the beam scan.}
\end{figure}

Numerical solutions of the heat diffusion equation are necessary to obtain the temporal and spatial dependence of temperature for a moving beam. As before, we neglect the small sideways heat transfer in the thin metallic film. One frame of the results is shown in figure 7 (the full movie is in the supplementary information). $F_{u}$ dominates at the center of the beam, where the gradients are small and temperatures the highest and, when the  conditions presented above are satisfied, pushes the DW from under the beam (figure 6). In addition, the heat front gives a force $F_{g}$ lined up with the temperature gradient (equation 6) and at an angle to the scanning direction. As the temperature decreases and $H_{c}$ increases, this transient heat wave front leaves its imprint in the magnetic structures observed with polarizing microscopy (figure 5). In contrast, calculations for substrates with $10\times D_{glass}$ and $\frac{D_{glass}}{10}$ diffusion constants give different heat profiles, not consistent with the polarization microscopy images (supplementary information). This confirms that it is the glass substrate, not the metal film, through which the heat transfer mostly occurs and shows the link between heat diffusion and domain wall motion predicted by the model presented.

To find the magnitudes of $F_{u,g}$ from equations 5 and 6, we need the time dependence of temperature $T(t)$ and heat fluxes $\overrightarrow{J}(t)=-k\overrightarrow{\nabla} T$. These are shown for point $A$ in the lower panels of figure 7. In addition, a quantitative calculation requires the $T$-dependence of $\sigma_{w}$ and $M_{s}$. These quantities cannot be measured in our setup at the high temperatures shown in figure 7. Estimates of $ F_{u,g}$, using typical $\sigma_{w}(T), M_{s}(T)$ dependencies for magnetic materials in equations 5 and 6, show values sufficient to overcome the coercive field of several tens of $G$ of sample $A$. The AOS observed in sample $B$, which had $H_{c}= 2.33~kG$ in the out-of-plane direction at room temperature, can be explained with a decrease of its coercive field at larger temperature. With these forces and a reduced coercive field from heat accumulation, the right-hand side in the velocity equation above becomes positive and the domain size increases. As the sample cools, the forces are reduced, while $H_{c}$ increases back to first stop the DW and, then, to prevent the expanded domains from collapsing. Therefore, the DW motion is not reversible and domains do not contract back on cooling once the laser beam moves away.

This model predicts that AOS is favored in samples with a high mobility $\eta$, a small $M_{s}$ and a small $H_{c}$ at large $T$. These predictions are consistent with observations. Sample $A$ had a smaller $H_c$ at room temperature and, in addition, a smaller thickness $h$. A smaller thickness would give a smaller magnetic anisotropy $K_{1}$ (this dependence was observed in Co/Pt~\cite{2016Hadri}), a larger DW width $\Delta\approx \pi \sqrt{\frac{A}{K_{1}}}$ and a larger Bloch DW mobility $\eta \approx\frac{\gamma \Delta}{\pi\alpha}$, where $\gamma, \alpha$ are the parameters in the LLG equation $\frac{d\overrightarrow{M}}{dt}=-\gamma \overrightarrow{M} \times \overrightarrow{H}_{eff}+\frac{\alpha}{M_{s}}\overrightarrow{M}\times \frac{d\overrightarrow{M}}{dt}$ rotating the magnetization at the DW location~\cite{1974Schryer}. It would be expected that AOS would be easier to observe in sample $A$. Indeed, AOS was obtained at lower powers and at the center of the stripe in sample $A$. Sample $B$ required higher powers and AOS was observed only at the stripe edges, where thermal gradients are larger. Previous measurements are also consistent with this model. For instance, it has been noted that AOS is favored at small $M_s$ and $M_{rem}$ in ferro- and ferrimagnetic materials. ~\cite{2016Hadri,2015Hassdenteufel}

The presented images were made long after AOS was complete and many questions remain, in particular on the domain nucleation process. Additional time-resolved experiments will address these questions.

\section{Conclusion}

All-optical switching by DW motion has been observed in ferromagnetic Co/Pd superlattices with high-repetition rate lasers. Once a domain is nucleated, a relatively slow domain evolution follows. Heat accumulation and in-plane diffusion are considered to obtain a force on the domain walls, which pushes the walls from under the laser spot.

\ack

This research was supported by the University of Louisville Research Foundation.

\vspace{50pt}

\underline{Supplementary material 1}

Time evolution of the temperature for a glass substrate (.avi video file). The time interval is $60~ms$. The parameters are the same as for figure 7 in the main text: $D_{glass}=0.005~\frac{cm^{2}}{s}$, $v_{scanning}=1~\frac{cm}{s}$, $F=5\times10^{7}~\frac{W}{m^{2}}$, and $A=30~\mu m \times 30~\mu m$. The plate size is $900~\mu m\times 180~\mu m\times 120~\mu m$. Starting temperature is $22^{o}~C$ and radiation losses are neglected.

\underline{Supplementary material 2}

Figure 8.

\begin{figure}
\centering\rotatebox{0}{\includegraphics[scale=0.5]{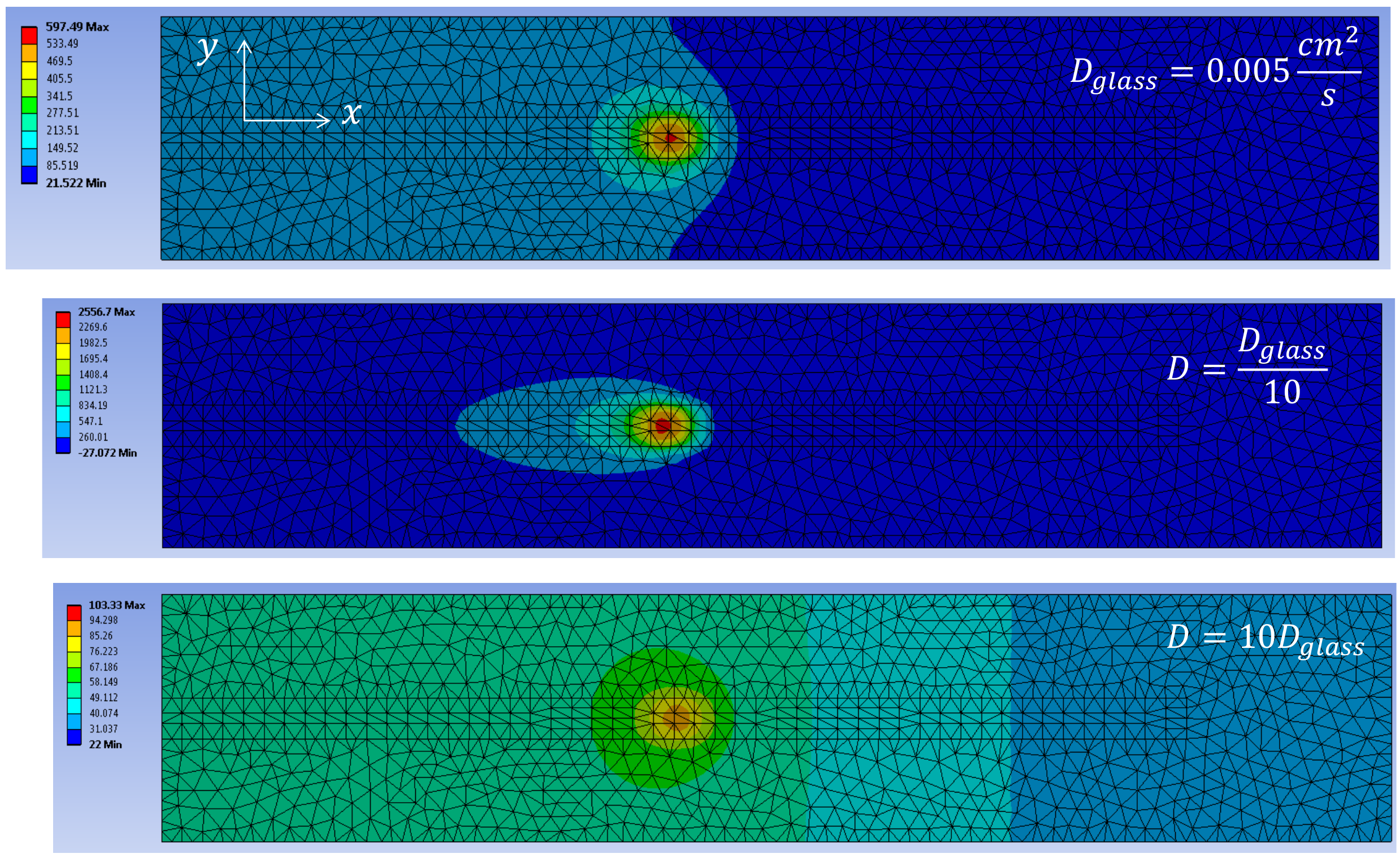}}
\caption{\label{fig:Figure1} Frames at the same time $t=37~ms$ for three different diffusion constants ($D_{glass}$, $\frac{D_{glass}}{10}$ and $10\times D_{glass}$). The peak temperature increases with lower $D$, as the heat accumulation is larger. The angle of the wave front is also different. This result can be understood qualitatively. For $D_{glass}=0.005~\frac{cm^{2}}{s}$, it takes $t\sim d^{2}/D=5~ms$ for heat to diffuse a distance $d=50~\mu m$. The beam has moved a distance $d_{s}=10~\frac{mm}{s}\times 5~ms=50 ~\mu m$ over this time. Therefore, the heat gradients in the substrate are on a time and length scale similar to that of the moving beam and the angle between the wave front normal and the scanning velocity is $\sim45^{0}$ in this case. Different diffusion constants and the same scanning speed result in different heat wave fronts, not consistent with polarizing microscopy images in figure 5.}
\end{figure}

\end{document}